\RequirePackage{fix-cm}
\documentclass[smallcondensed]{svjour3}     
\smartqed  
\usepackage{graphicx}
\usepackage{natbib}
\usepackage{amsmath, amssymb}
\usepackage{algorithm}
\usepackage{algorithmic}
\usepackage{tikz}
\usepackage[linkcolor=blue,citecolor=blue,colorlinks]{hyperref}
\newcommand{\e}{\mathrm{e}}
\newcommand{\ii}{\mathrm{i}}
\newcommand{\trans}[1]{{#1}^T}
\newcommand{\conj}[1]{{\bar{#1}}}

\newcommand{\field}[1]{\cal {#1}}
\newcommand{\hfield}[1]{\hat{\field {#1}}}
\newcommand{\hfvec}[1]{\hat{\vec{\field {#1}}}}
\newcommand{\mat}[1]{\vec{#1}}
\newcommand{\dd}{\mathrm{d}}

\defcitealias{Hou16}{HSX16}
\begin{document}

\title{The Two Rigid Body Interaction using Angular Momentum Theory Formulae}

\author{Gwena\"el Bou\'e}

\authorrunning{G. Bou\'e} 

\institute{G. Bou\'e  \at
              \email{gwenael.boue@obspm.fr}
              IMCCE, Observatoire de Paris, UPMC Univ. Paris 6, Paris, France
}

\date{Received: date / Accepted: date}

\graphicspath{ {figures/} }

\maketitle

\begin{abstract}
This work presents an elegant formalism to model the evolution of the
full two rigid body problem. The equations of motion, given in a
Cartesian coordinate system, are expressed in terms of spherical
harmonics and Wigner D-matrices. The algorithm benefits from the
numerous recurrence relations satisfied by these functions allowing a
fast evaluation of the mutual potential. Moreover, forces and
torques are straightforwardly obtained by application of ladder
operators taken from the angular momentum theory and commonly used in
quantum mechanics. A numerical implementation of this algorithm is
made. Tests show that the present code is significantly faster than
those currently available in literature.

\end{abstract}


\section{Introduction}
\label{intro}
Modelling the evolution of two rigid bodies with arbitrary shapes in
gravitational interaction is not an easy task. The first obstacle is the
determination of the mutual potential, the second is the derivation of
the equations of motion in a suitable form to allow fast computation.
In general cases, the potential has to be expanded and truncated at some
order in the ratio of the bodies mean radii to the distance between the two
barycenters. Using the angular momentum theory developed by
\citet{Wigner59}, \citet{Borderies78} manage to provide a compact
expression of the mutual potential at any order. In this expression, the
gravity field of each body is described by Stokes coefficients, the
relative orientation of the two bodies appears through Wigner D-matrices
-- also called Euler functions \citep{Borderies78} -- and the
dependence in the distance between the two barycenters is embedded in
solid harmonics. Equations of motion associated to this expansion have
been proposed by \citet{Maciejewski95}. A strictly equivalent formalism
based on symmetric trace-free (STF) tensor \citep{Hartmann94, Mathis09}
has been implemented by \citet{Compere14} who studied numerically the
evolution of the binary asteroid 1999 KW4.

The decomposition in spherical harmonics has sometimes been discarded
based on the misconception that this formalism involves many
trigonometric functions increasing as much the computation time and the
risk of numerical instabilities. To circumvent this issue, alternative
decompositions of the potential have been applied. For instance,
\citet{Paul88} explicitly wrote the mutual potential in Cartesian
coordinates at all orders. The associated equations of motion were then
provided by \citet{Tricarico08}. Recently, \citet{Hou16} (hereafter,
\citetalias{Hou16}) revisited this approach and built the as yet fastest
algorithm able to integrate the full two rigid body problem thanks to a
set of recurrence formulae whose coefficients can be computed and stored
beforehand. Another example of alternative is the so-called polyhedron
approach relying on Stokes' theorem in which the volume integral leading
to the mutual potential is converted into a surface integral over the
boundaries of the two bodies \citep{Werner05}. This algorithm
implemented by \citet{Fahnestock06} was at the time the fastest
integrator.

Here we revisit \citeauthor{Borderies78}' expression of the mutual
potential developed in spherical harmonics, but we express them in
Cartesian coordinates. By consequence, the description of the orbital
configuration is actually equivalent to those where the potential is
explicitly written in Cartesian coordinates as in \citep{Paul88}, for
instance. The main contribution of the present study is in the
parametrisation of rotations. Whereas in previous works, torques are
computed by an explicit derivation of the potential energy with respect
to Euler angles or with respect to rotation matrix elements, here forces
and torques are simply obtained by application of ladder operators
taken from quantum mechanics theory. This approach has already been
successfully applied to the modelling of tidal evolution of close-in
planets \citep{Boue16}. In this study we show that it allows to built
the fastest numerical integrator of the full two rigid body problem.

The paper is organised as follows: in Section~\ref{sec.eom}, the
equations of motion are computed using Poincar\'e's method
\citep{Poincare01}. Forces and torques are written in terms of ladder
operators commonly used in quantum mechanics. The section also contains
all the recurrence relations allowing an efficient numerical
implementation of the problem.  Numerical tests are performed in
Section~\ref{sec.numerical}. Conclusions are drawn in the last section.

\section{Equations of motion}
\label{sec.eom}
Let two rigid bodies $A$ and $B$ with arbitrary shapes in gravitational
interaction in an inertial frame $(O; \vec e_x, \vec e_y, \vec e_z)$.
For both bodies $A$ and $B$ we define body-fixed frames centred on
their respective barycenters $O_A$ and $O_B$ which are $(O_A; \vec
e^A_x, \vec e^A_y, \vec e^A_z)$ and $(O_B; \vec e^B_x, \vec e^B_y, \vec
e^B_z)$, respectively.
To fasten the integration, the problem is not described in the inertial
frame as in \citep{Borderies78}, but in the body-fixed frame of $A$
\citep{Maciejewski95}. This is also the choice of \citetalias{Hou16} with
whom we wish to compare the method. The convention used in this paper is the
following: unprimed quantities are expressed in the body-fixed frame of
$A$, while primed ones are written in the body-fixed frame of $B$.
Vectors expressed in the inertial frame are written with the superscript
$0$.
\subsection{Lagrangian}
Following \citet{Maciejewski95}, the potential reads
\begin{equation}
\begin{split}
U_{AB} =& - GM_AM_B 
       \sum_{l_1=0}^\infty \sum_{m_1=-l_1}^{l_1}
       \sum_{l_2=0}^\infty \sum_{m_2=-l_2}^{l_2}
       \bigg(
       R_A^{l_1} R_B^{l_2}
       Z_{l_1,m_1}^A Z^B_{l_2,m_2}
\\ & \times
       (-1)^{l_2} \gamma^{l_1,m_1}_{l_2,m_2}
       \frac{Y_{l_1+l_2,m_1+m_2}(\vec r)}{r^{l_1+l_2+1}}
       \bigg)
\end{split}
\label{eq.Uab}
\end{equation}
where $M_*$, $R_*$, and $Z^*_{l,m}$ are the mass, the mean radius and
Stokes coefficients of the body $*=A,B$, respectively. Note that
$Z^A_{l,m}$ is constant while $Z^B_{l,m}$ is not because both quantities
are written in the body-fixed frame of $A$. The radius vector $\vec r =
\vec r_B - \vec r_A$ connects the two barycenters. $Y_{l,m}$ are
complex spherical harmonics and $\gamma^{l_1,m_1}_{l_2,m_2}$ are
constant coefficients.  Here we use the Schmidt semi-normalisation of
the spherical harmonics, such that,
\begin{equation}
\label{eq.Ylm}
Y_{l,m}(\theta, \phi) = (-1)^m \sqrt{\frac{(l-m)!}{(l+m)!}} P_{l,m}(\cos
\theta) \e^{\ii m \phi}
\end{equation}
where the associated Legendre polynomials $P_{l,m}$ are defined as
$$
P_{l,m}(x) =
\frac{1}{2^ll!}(1-x^2)^{m/2}\frac{\dd^{l+m}}{\dd x^{l+m}}(x^2-1)^l.
$$
The complex Stokes coefficients $Z_{l,m}$ are related to
the usual real Stokes coefficients $C_{l,m}$ and $S_{l,m}$ by
\begin{subequations}
\begin{equation}
Z_{l,m} = (-1)^m \frac{1+\delta_{m,0}}{2} \sqrt{\frac{(l+m)!}{(l-m)!}}
(C_{l,m} - \ii S_{l,m})\ , \quad\mathrm{for}\quad m\geq 0
\label{eq.Zlm}
\end{equation}
and the symmetry relation
\begin{equation}
\label{eq.symZlm}
Z_{l,m} = (-1)^m \conj Z_{l,-m} \ ,\quad\mathrm{for}\quad m<0\ .
\end{equation}
\end{subequations}
In Eq.~(\ref{eq.Zlm}), $\delta_{i,j}$ is the Kronecker delta equal to 1
if $i=j$ and equal to 0 otherwise. In Eq.~(\ref{eq.symZlm}), the bar
above $Z_{l,m}$ denotes the complex conjugate. Following our
normalisation convention, the coefficients $\gamma^{l_1,m_1}_{l_2,m_2}$
are
$$
\gamma^{l_1,m_1}_{l_2,m_2} = 
  \sqrt{\frac{(l_1+l_2-m_1-m_2)!(l_1+l_2+m_1+m_2)!}
        {(l_1+m_1)!(l_1-m_1)!(l_2+m_2)!(l_2-m_2)!}}.
$$
Let
$$
\mat C_A = [\vec e^A_x, \vec e^A_y, \vec e^A_z],
\qquad
\mat C_B = [\vec e^B_x, \vec e^B_y, \vec e^B_z]
\qquad\mathrm{and}\qquad
\mat C = \trans{\mat C}_A \mat C_B
$$
be the rotation matrices such that
$$
\vec r = \trans{\mat C}_A \vec r^0,
\qquad
Z^B_{l,m} = \sum_{m'=-l}^l D^{l}_{m,m'}(\mat C) Z'^B_{l,m'}\ ,
$$
where $\trans{\mat C}_A$ means the transpose of $\mat C_A$ and
$D^l_{m,m'}(\mat C)$ is an element of Wigner D-matrix associated to the
rotation $\mat C$. $Z'^B_{l,m'}$ are the (constant) complex Stokes
coefficients of body $B$ expressed in its body-fixed frame.

In the barycentric frame, the kinetic energy of the system reads
\begin{equation}
T = \frac{1}{2}\trans{\vec \Omega}_A \mat I_A \vec \Omega_A
   +\frac{1}{2}\trans{(\vec \Omega + \vec \Omega_A)} \mat I_B
               (\vec \Omega + \vec \Omega_A)
   +\frac{1}{2} \mu \|\vec v + \vec \Omega_A\times \vec r\|^2
\label{eq.kinetic}
\end{equation}
where $\mat I_A$ and $\mat I_B$ are the inertia matrices of the bodies $A$
and $B$ expressed in the the body-fixed frame of $A$, respectively.
$\vec \Omega_A$ is the rotation vector of the body $A$ with respect to the
inertial frame, while $\vec \Omega = \vec \Omega_B - \vec \Omega_A$ is
the rotation vector of the body $B$ relative to the body $A$.
$\mu=M_AM_B/(M_A+M_B)$ is the reduced mass and $\vec v = \dot{\vec r}$
is the relative velocity in the body-fixed frame of $A$. In
the expression of the kinetic energy (Eq.~\ref{eq.kinetic}), $\mat I_A$
is constant while $\mat I_B = \mat C \mat I'_B \trans{\mat C}$ is a
function of $\mat C$.

The Lagrangian of the problem $f = T - U_{AB}$ is a function of $\vec q
= (\mat C_A, \mat C, \vec r)$ and $\vec \eta = (\vec \Omega_A, \vec
\Omega, \vec v)$. Note that $\mat C_A$ and $\mat C$ are themselves
functions of only three coordinates, such as the 3-2-3 Euler angles
$(\alpha_A, \beta_A, \gamma_A)$ and $(\alpha, \beta, \gamma)$,
respectively:
$$
\mat C_A(\alpha_A, \beta_A, \gamma_A) = 
  \mat C_z(\alpha_A)\mat C_y(\beta_A)\mat C_z(\gamma_A),
\qquad
\mat C(\alpha, \beta, \gamma) = 
  \mat C_z(\alpha)\mat C_y(\beta)\mat C_z(\gamma).
$$
To retrieve the equations of motion given by \citet{Maciejewski95}, we
use Poincar\'e's {\em forme nouvelle des \'equations de la m\'ecanique}
\citep{Poincare01}.  We define a vector field basis associated to
infinitesimal rotations of the body $A$ as
\begin{subequations}
\begin{eqnarray}
\hfield J_{A,x} &=& 
   \cos\gamma_A\cot\beta_A\frac{\partial}{\partial \gamma_A}
  +\sin\gamma_A\frac{\partial}{\partial \beta_A}
  -\frac{\cos\gamma_A}{\sin\beta_A}\frac{\partial}{\partial \alpha_A},
\\
\hfield J_{A,y} &=&
  -\sin\gamma_A\cot\beta_A\frac{\partial}{\partial \gamma_A}
  +\cos\gamma_A\frac{\partial}{\partial \beta_A}
  +\frac{\sin\gamma_A}{\sin\beta_A}\frac{\partial}{\partial \alpha_A},
\\
\hfield J_{A,z} &=& 
   \frac{\partial}{\partial \gamma_A}.
\end{eqnarray}%
\label{eq.JA}%
\end{subequations}
Equivalently, we introduce a basis field corresponding to infinitesimal
rotations of the body $B$ relative to the body $A$: 
\begin{subequations}
\begin{eqnarray}
\hfield J_{x} &=&
  -\cos\alpha\cot\beta\frac{\partial}{\partial \alpha}
  -\sin\alpha\frac{\partial}{\partial \beta}
  +\frac{\cos\alpha}{\sin\beta}\frac{\partial}{\partial \gamma},
\\
\hfield J_{y} &=&
  -\sin\alpha\cot\beta\frac{\partial}{\partial \alpha}
  +\cos\alpha\frac{\partial}{\partial \beta}
  +\frac{\sin\alpha}{\sin\beta}\frac{\partial}{\partial \gamma},
\\
\hfield J_{z} &=&
   \frac{\partial}{\partial \alpha}.
\end{eqnarray}%
\label{eq.J}%
\end{subequations}
At last, we consider the canonical vector field basis associated to
infinitesimal translations
\begin{subequations}
\begin{eqnarray}
\hfield P_{x} &=& \frac{\partial}{\partial x}, \\
\hfield P_{y} &=& \frac{\partial}{\partial y}, \\
\hfield P_{z} &=& \frac{\partial}{\partial z}.
\end{eqnarray}%
\label{eq.P}%
\end{subequations}
Let us gather all these vector fields into a single vector $\hfvec X =
(\hfvec J_A, \hfvec J, \hfvec P)$. The generalised velocity vector
$\dot{\vec q}$ of the Lagrangian reads
\begin{equation}
\dot{\vec q} = \vec \eta \cdot \hfvec X(\vec q) \equiv
\sum_{i=1}^9 \eta_i \hfield X_i(\vec q),
\label{eq.dotq}
\end{equation}
as required by \citeauthor{Poincare01}'s formalism.  
The equations of motion of $\mat C_A$, $\mat C$, and $\vec r$
given by Eq.~(\ref{eq.dotq}) are made explicit in the following.
To obtain the equations of motion, we need the structure constants
$c_{ij}^k$ defined as
$$
[\hfield X_i, \hfield X_j] \equiv \hfield X_i \hfield X_j - \hfield X_j \hfield X_i = c_{ij}^k \hfield X_k.
$$
A direct calculation shows that the non-vanishing commutators are
\begin{equation}
\begin{array}{ll}
&[\hfield J_{A,x}, \hfield J_{A,y}] = -\hfield J_{A,z}, \\[0.5em]
&[\hfield J_{A,y}, \hfield J_{A,z}] = -\hfield J_{A,x}, \\[0.5em]
&[\hfield J_{A,z}, \hfield J_{A,x}] = -\hfield J_{A,y}, 
\end{array}
\qquad
\begin{array}{ll}
&[\hfield J_{x}, \hfield J_{y}] = \hfield J_{z}, \\[0.5em]
&[\hfield J_{y}, \hfield J_{z}] = \hfield J_{x}, \\[0.5em]
&[\hfield J_{z}, \hfield J_{x}] = \hfield J_{y}.
\end{array}
\label{eq.commutator}
\end{equation}
To get \citeauthor{Poincare01}'s equation, we also need to evaluate
$\hfield X_i(f)$. We have
\begin{subequations}
\begin{eqnarray}
\label{eq.JAf}
\hfvec J_A(f) &=& \vec 0, \\[0.5em]
\label{eq.Jf}
\hfvec J(f) &=& \vec G_B \times (\vec \Omega + \vec \Omega_A) - \hfvec J(U_{AB}), \\[0.5em]
\label{eq.Pf}
\hfvec P(f) &=& \mu \vec V \times \vec \Omega_A - \hfvec P(U_{AB}).
\end{eqnarray}%
\label{eq.Xf}%
\end{subequations}
The equation~(\ref{eq.JAf}) results from the invariance by rotation of
the problem. Indeed, the Lagrangian does not depend on $\mat C_A$. In
Eq.~(\ref{eq.Pf}), $\vec V = \vec v + \vec \Omega_A \times \vec r$
represents the velocity relative to the barycentric frame expressed in
the body-fixed frame of $A$. In Eq.~(\ref{eq.Jf}), we have introduce
$\vec G_B = \mat I_B (\vec \Omega + \vec \Omega_A)$, the angular
momentum of the body $B$. Similarly, we denote by $\vec G_A = \mat I_A
\vec \Omega_A$ the angular momentum of the body $A$, and by $\vec L =
\mu \vec r \times \vec V$, the orbital angular momentum. The last
ingredients in \citeauthor{Poincare01}'s equations are the partial
derivatives of the kinetic energy with respect to $\vec \eta$ which are
\begin{equation}
\frac{\partial T}{\partial \vec \Omega_A} = \vec G_A + \vec G_B + \vec L, 
\qquad
\frac{\partial T}{\partial \vec \Omega} = \vec G_B,
\qquad
\frac{\partial T}{\partial \vec v} = \mu \vec V.
\label{eq.dTdeta}
\end{equation}
Combining Eqs.~(\ref{eq.dotq}), (\ref{eq.commutator}), (\ref{eq.Xf}),
(\ref{eq.dTdeta}), and \citeauthor{Poincare01}'s equations
$$
\frac{d}{dt}\frac{\partial T}{\partial \eta_i} = \sum_{j,k} c_{ij}^k
\eta_j \frac{\partial T}{\partial \eta_k} + \hfield X_i(f),
$$
we get
\begin{equation}
\left\{
\begin{array}{ll}
\dot{\vec C}_A = \vec \Omega_A \cdot \hfvec J_A(\mat C_A), \\[0.5em]
\dot{\vec C} = (\vec \Omega_B-\vec\Omega_A)\cdot\hfvec J(\mat C), \\[0.5em]
\dot{\vec r} = \vec r\times\vec\Omega_A + \vec V,
\end{array}
\right.
\qquad
\left\{
\begin{array}{ll}
\dot{\vec G}_A = \vec G_A\times\vec\Omega_A + \vec T_A, \\[0.5em]
\dot{\vec G}_B = \vec G_B\times\vec\Omega_A + \vec T_B, \\[0.5em]
\dot{\vec V}   = \vec V \times \vec \Omega_A+ \frac{1}{\mu}\vec F,
\end{array}
\right.
\label{eom}
\end{equation}
with
$$
\vec F   = -\hfvec P(U_{AB}),
\qquad
\vec T   = -\hfvec L(U_{AB}),
\qquad
\vec T_B = -\hfvec J(U_{AB}),
\qquad
\vec T_A = -\vec T - \vec T_B.
$$
The rotation vectors are given by
$$
\vec \Omega_A = \mat I_A^{-1} \vec G_A,
\qquad
\vec \Omega_B = \mat I^{-1}_B \vec G_B,
$$
and the vector field $\hfvec L$ is defined as
\begin{subequations}
\begin{eqnarray}
\hfield L_x &=& y\frac{\partial }{\partial z} - z\frac{\partial }{\partial y}, \\[0.5em]
\hfield L_y &=& z\frac{\partial }{\partial x} - x\frac{\partial }{\partial z}, \\[0.5em]
\hfield L_z &=& x\frac{\partial }{\partial y} - y\frac{\partial }{\partial x}.
\end{eqnarray}%
\label{eq.L}%
\end{subequations}
The equations (\ref{eom}) are strictly equivalent to
\citeauthor{Maciejewski95}'s equations of motion, but they are
explicitly written in terms of the vector fields $\hfvec P$, $\hfvec L$,
$\hfvec J_A$ and $\hfvec J$. As shown in the subsequent section,
textbooks on quantum theory of angular momentum allow to evaluate these
vector fields applied to $U_{AB}$ very efficiently.

\subsection{Ladder operators, force and torques}
\label{sec.ladder}
We introduce a complex coordinate system $(\vec e_+, \vec e_0, \vec
e_-)$ in which coordinates of any vector $\vec a$ are denoted $(a_+,
a_0, a_-)$ and are related to the usual Cartesian coordinates $(a_x,
a_y, a_z)$ by
$$
a_+ = -\frac{1}{\sqrt 2}(a_x + \ii a_y)
\qquad
a_0 = a_z
\qquad
a_- = \frac{1}{\sqrt 2}(a_x - \ii a_y)\ .
$$
We apply the same rule for vector fields $\hfvec A = (\hfield A_x,
\hfield A_y,
\hfield A_z)$. We get
$$
\hfield A_+ = -\frac{1}{\sqrt 2}(\hfield A_x + \ii \hfield A_y)
\qquad
\hfield A_0 = \hfield A_z
\qquad
\hfield A_- = \frac{1}{\sqrt 2}(\hfield A_x - \ii \hfield A_y).
$$
The vector fields defined in Eqs.~(\ref{eq.JA}), (\ref{eq.J}),
(\ref{eq.P}), and (\ref{eq.L}) are related to ladder operators commonly
used in quantum theory of angular momentum. With the notation of
\citet{Varshalovich88}, we have
$$
\hfield J_{A,\nu} \equiv \ii\hat J'^\nu(\alpha_A, \beta_A, \gamma_A),
\qquad
\hfield J_\nu \equiv \ii\hat J_{\nu}(\alpha, \beta, \gamma),
\qquad
\hfield P_\nu \equiv \nabla_\nu,
\qquad
\hfield L_\nu \equiv \ii\hat L_\nu,
$$
for $\nu\in\{+,0,-\}$. $\hat J'^\nu$ and $\hat J_\nu$ are the components
of the spin operator expressed in the rotated frame and in the initial
frame, respectively. $\nabla_\nu$ is the usual gradient operator and
$\hat L_\nu$ is the orbital angular momentum operator. Given these
equivalences, we deduce the following relations \citep{Varshalovich88}
\begin{subequations}
\begin{equation}
\hfield P_\nu \left(\frac{Y_{l,m}(\vec r)}{r^{l+1}}\right) = 
\left\{
\begin{array}{ll}
\displaystyle-\sqrt{(l+m+1)(l-m+1)}\frac{Y_{l+1,m}(\vec r)}{r^{l+2}}\qquad &\nu=0 \\[0.5em]
\displaystyle-\sqrt{\frac{(l\pm m+1)(l\pm m+2)}{2}} \frac{Y_{l+1,m\pm1}(\vec r)}{r^{l+2}}\quad &\nu=\pm1
\end{array}
\right.
\label{eq.Ladder1}
\end{equation}
and
\begin{equation}
\hfield L_\nu \left(\frac{Y_{l,m}(\vec r)}{r^{l+1}}\right) = \ii 
\left\{
\begin{array}{ll}
\displaystyle m \frac{Y_{l,m}(\vec r)}{r^{l+1}}\qquad &\nu=0 \\
\displaystyle \mp
\sqrt{\frac{l(l+1)-m(m\pm1)}{2}}\frac{Y_{l,m\pm1}(\vec r)}{r^{l+1}}\quad
&\nu=\pm1
\end{array}
\right.
\ ,
\label{eq.Ladder2}
\end{equation}%
\label{eq.Ladder12}%
\end{subequations}
for the orbital part (embedded in the spherical harmonics), while spin
operators act on Wigner D-matrices according to
\begin{subequations}
\begin{equation}
\hfield J_\nu D^l_{m,m'}(\mat C) = \ii
\left\{
\begin{array}{ll}
-m D^l_{m,m'}(\mat C)\ ,\qquad & \nu=0, \\[0.5em]
\pm\sqrt{\displaystyle\frac{l(l+1)-m(m\mp1)}{2}}D^l_{m\mp1,m'}(\mat C)\,\quad & \nu=\pm1
\end{array}
\right.
\label{eq.Ladder3}
\end{equation}
and
\begin{equation}
\hfield J_{A,\nu} D^l_{m,m'}(\mat C_A) = \ii
\left\{
\begin{array}{ll}
-m' D^l_{m,m'}(\mat C_A)\ ,\qquad & \nu=0, \\[0.5em]
\pm\sqrt{\displaystyle\frac{l(l+1)-m'(m'\pm1)}{2}}D^l_{m,m'\pm1}(\mat C_A)\,\quad & \nu=\pm1
\end{array}
\right.\ .
\label{eq.J'}
\end{equation}%
\label{eq.Ladder34}%
\end{subequations}

Forces and torques are then computed as follows.  Let us define the
constant terms
$u^{l_1,m_1}_{l_2,m_2}$ as
\begin{equation}
u^{l_1,m_1}_{l_2,m_2} = -GM_AM_BR_A^{l_1}R_B^{l_2}Z^A_{l_1,m_1}
                         (-1)^{l_2}\gamma^{l_1,m_1}_{l_2,m_2}.
\label{eq.u}
\end{equation}
The potential (Eq.~\ref{eq.Uab}) reads
$$
U_{AB} =
 \sum_{l_1=0}^\infty \sum_{m_1=-l_1}^{l_1}
 \sum_{l_2=0}^\infty \sum_{m_2=-l_2}^{l_2}
 u^{l_1,m_1}_{l_2,m_2} Z^B_{l_2,m_2}(\mat C)
 \frac{Y_{l_1+l_2,m_1+m_2}(\vec r)}{r^{l_1+l_2+1}},
$$
and thus,
\begin{align*}
&\vec F =
 -
 \sum_{l_1=0}^\infty \sum_{m_1=-l_1}^{l_1}
 \sum_{l_2=0}^\infty \sum_{m_2=-l_2}^{l_2}
 u^{l_1,m_1}_{l_2,m_2} Z^B_{l_2,m_2}(\mat C)
 \hfvec P\left(\frac{Y_{l_1+l_2,m_1+m_2}(\vec r)}{r^{l_1+l_2+1}}\right),
\\
&\vec T =
 -
 \sum_{l_1=0}^\infty \sum_{m_1=-l_1}^{l_1}
 \sum_{l_2=0}^\infty \sum_{m_2=-l_2}^{l_2}
 u^{l_1,m_1}_{l_2,m_2} Z^B_{l_2,m_2}(\mat C)
 \hfvec L\left(\frac{Y_{l_1+l_2,m_1+m_2}(\vec r)}{r^{l_1+l_2+1}}\right),
\\
&\vec T_B =
 -
 \sum_{l_1=0}^\infty \sum_{m_1=-l_1}^{l_1}
 \sum_{l_2=0}^\infty \sum_{m_2=-l_2}^{l_2}
 u^{l_1,m_1}_{l_2,m_2} 
 \hfvec J \left( Z^B_{l_2,m_2}(\mat C) \right)
 \frac{Y_{l_1+l_2,m_1+m_2}(\vec r)}{r^{l_1+l_2+1}},
\\
&\vec T_A = -\vec T - \vec T_B\ ,
\end{align*}
with
$$
Z^B_{l_2,m_2}(\mat C) = \sum_{m'_2=-l_2}^{l_2} D^{l_2}_{m_2,m'_2}(\mat C) Z'^B_{l_2,m'_2},
$$
and
$$
\hfvec J\left(Z^B_{l_2,m_2}(\mat C)\right) = \sum_{m'_2=-l_2}^{l_2} \hfvec
J \left(D^{l_2}_{m_2,m'_2}(\mat C)\right) Z'^B_{l_2,m'_2}.
$$
The simplicity of the formulae (\ref{eq.Ladder12}) and
(\ref{eq.Ladder34}) associated to the evaluation of the ladder operators
makes the calculation of the components of $\vec F$, $\vec T$, and $\vec T_B$
very efficient and easy to implement.

\subsection{Spherical harmonics}
Naturally, spherical harmonics are not computed from their definition
Eq.~(\ref{eq.Ylm}) but from recurrence formulae which can also be found
in many textbooks. Let $\vec u \equiv \vec r/\|\vec r\| = (u_+,u_0,u_-)$
be the unit vector along $\vec r$. By definition $Y_{l,m}(\vec r)
\equiv Y_{l,m}(\vec u)$.  The initialisation is
\begin{subequations}
\begin{equation}
Y_{0,0}(\vec r) = 1,  \quad
Y_{1,0}(\vec r) = u_0, \quad
Y_{1,1}(\vec r) = u_+,
\end{equation}
and the recurrence equations are
\begin{align}
\label{eq.Ylm1}
lY_{l,0}(\vec r) &= (2l-1)u_0Y_{l-1,0}(\vec r)-(l-1)Y_{l-2,0}(\vec r), \\
\label{eq.Ylm2}
\sqrt{l+m}Y_{l,m}(\vec r) &= \sqrt{l-m}\,u_0Y_{l-1,m}(\vec r)+\sqrt{2(l+m-1)}\,u_+Y_{l-1,m-1}(\vec r)\ .
\end{align}
Spherical harmonics with negative order $m$ are deduced from the symmetry
relation
\begin{equation}
Y_{l,m}(\vec r) = (-1)^m \conj Y_{l,-m}(\vec r)\ .
\end{equation}%
\label{eqs.Ylm}%
\end{subequations}
As it can be noticed, in the set of equations (\ref{eqs.Ylm}), there
isn't any trigonometric functions.

\subsection{Cayley-Klein parameters}
To avoid singularities, rotations -- involved in Wigner D-matrices --
are para\-me\-trised by Cayley-Klein complex parameters $a$ and $b$
\citep[e.g.,][]{Varshalovich88} instead of the more common Euler 3-1-3
angles $(\psi, \theta, \phi)$ or the Euler 3-2-3 angles
$(\alpha,\beta,\gamma)$. Cayley-Klein parameters are equivalent to
quaternions, but are more adapted to the formalism used in this work.
For the sake of completeness, we here summarise a few properties of
these parameters \citep{Varshalovich88}. They and Euler 3-1-3 angles are
related to each others by
\begin{subequations}
\begin{equation}
a = \cos\frac{\theta}{2} \e^{-\ii\frac{\psi+\phi}{2}} 
  = D^{\frac{1}{2}}_{\frac{1}{2},\frac{1}{2}}(\mat C)\ ,
\qquad
b = -\ii \sin\frac{\theta}{2} \e^{\ii\frac{\psi-\phi}{2}}
  = D^{\frac{1}{2}}_{-\frac{1}{2},\frac{1}{2}}(\mat C)\ ,
\label{eq.defab}
\end{equation}
and reciprocally,
\begin{equation}
\psi   = \arg(\ii \conj{a}b)\ ,
\qquad
\theta = \cos^{-1}\left(|a|^2-|b|^2\right)\ ,
\qquad
\phi   = -\arg(\ii a b)\ .
\end{equation}
\end{subequations}
The expression of the Cartesian coordinates of a rotated vector, e.g.
$\vec r = \mat C \vec r'$, expressed in terms of Cayley-Klein parameters is
$$
r_0 = \left(|a|^2-|b|^2\right)r'_0 + 2\sqrt{2}\,\Re\!\left(ab\,\conj r'_+\right)
\qquad
r_+ = -\sqrt 2 \conj a b\,r'_0 + \conj a^2 r'_+ - b^2 \conj r'_+\ ,
$$
where $\Re(z)$ means the real part of $z$.
The inverse rotation is obtained by the substitution $(a,b) \rightarrow
(\conj a, -b)$. The product of two rotations $\mat C(a,b) = \mat C(a_1,b_1)
\mat C(a_2,b_2)$ is given by
$$
a = a_1 a_2 - \conj b_1 b_2\ ,
\qquad
b = b_1 a_2 + \conj a_1 b_2\ .
$$
Finally, the equivalent of \citeauthor{Maciejewski95}'s equations of
motion of $\mat C$ and $\mat C_A$ (Eq.~\ref{eom}) in terms of
Cayley-Klein parameters ($a,b$) are deduced from their expressions in terms of
Wigner D-matrix (Eq.~\ref{eq.defab}). Indeed, for all $(l,m,m')$, we
have
$$
\left\{
\begin{array}{ll}
\dot D^l_{m,m'}(\mat C)   &=  (\vec \Omega_B - \vec \Omega_A) \cdot
\hfvec J\left( D^l_{m,m'}(\mat C)\right) \\[1.2em]
\dot D^l_{m,m'}(\mat C_A) &= \vec \Omega_A \cdot \hfvec J_A \left(D^l_{m,m'}(\mat C_A)\right)
\end{array}
\right. .
$$
Taking the particular values of $(l,m,m')$ corresponding to the
definitions of $a$ and $b$ (Eq.~\ref{eq.defab}), we get
\begin{subequations}
\begin{equation}
\left\{
\begin{array}{ll}
\dot a &= -\displaystyle
          \frac{\ii}{2}\left(a(\Omega_{B,0} -\Omega_{A,0}) -
          \sqrt{2}b(\conj \Omega_{B,+} - \conj\Omega_{A,+})\right)
\\[1.2em]
\dot b &= \displaystyle
         +\frac{\ii}{2}\left(b(\Omega_{B,0} -\Omega_{A,0}) +
          \sqrt{2}a(\Omega_{B,+} - \Omega_{A,+})\right)
\end{array}
\right.
\end{equation}
and
\begin{equation}
\left\{
\begin{array}{ll}
\dot a_A &= -\displaystyle
          \frac{\ii}{2}\left(a_A\Omega_{A,0} + 
          \sqrt{2}\,\conj{b}_A\Omega_{A,+}\right)
\\[1.2em]
\dot b_A &= \displaystyle
         -\frac{\ii}{2}\left(b_A\Omega_{A,0} -
          \sqrt{2}\,\conj{a}_A\Omega_{A,+}\right)
\end{array}
\right. .
\end{equation}
\label{eom.ab}
\end{subequations}

\subsection{Wigner D-matrices}
Like for spherical harmonics, the computation of Wigner D-matrices
is performed recursively. Here, we need these matrices with
integer $l$. The recurrence is initialised with
$$
D^0_{0,0} = 1\ ,
\qquad
D^1_{m,m'} = \begin{array}{cc}
{\footnotesize (1\hspace{1.1cm} 0\hspace{1.1cm}-1)} \\[0.2em]
\arraycolsep=3.5pt\def\arraystretch{1.6}
\begin{bmatrix}
a^2 & -\sqrt{2} a\conj{b} & (\conj{b})^2 \\
\sqrt{2} ab & |a|^2-|b|^2 & -\sqrt{2}\conj{a}\conj{b} \\
b^2 & \sqrt{2}\conj{a}b & (\conj{a})^2
\end{bmatrix} & 
\rotatebox[origin=c]{-90}{(1\hspace{0.3cm} 0\hspace{0.3cm} -1)} \\
\color{white}{0}
\end{array}\ .
\vspace{-1em}  
$$
For $m\geq 0$, Wigner D-matrices of degree $l\geq 2$ are given by
\citep[see][]{Gimbutas09}
\begin{subequations}
\begin{equation}
D^l_{m,m'} = c^{l,-}_{m,m'} D^1_{1,1}D^{l-1}_{m-1,m'-1}
           + c^{l,0}_{m,m'} D^1_{1,0}D^{l-1}_{m-1,m'}
           + c^{l,+}_{m,m'} D^1_{1,-1}D^{l-1}_{m-1,m'+1}
\label{eq.recD}
\end{equation}
with coefficients
\begin{equation}
\left\{
\begin{array}{l}
\displaystyle c^{l,-}_{m,m'} = \sqrt{\frac{(l+m')(l+m'-1)}{(l+m)(l+m-1)}}, \\[1.2em]
\displaystyle c^{l,0}_{m,m'} = \sqrt{\frac{2(l+m')(l-m')}{(l+m)(l+m-1)}}, \\[1.2em]
\displaystyle c^{l,+}_{m,m'} = \sqrt{\frac{(l-m')(l-m'-1)}{(l+m)(l+m-1)}}\ .
\end{array}
\right.
\label{eq.recD2}
\end{equation}
\end{subequations}
If $|m'+\nu|$, with $\nu\in\{-1,0,1\}$, is
strictly greater than $l-1$, then $D^{l-1}_{m-1,m'+\nu}$ should be
discarded and replaced by zero in the left-hand side of Eq.~(\ref{eq.recD}). Elements of
Wigner D-matrices with negative index $m$ are deduced from the symmetry
relation
$$
D^l_{m,m'} = (-1)^{m-m'} \conj D^l_{-m,-m'}\ .
$$

\section{Numerical implementation}
\label{sec.numerical}
The algorithm presented above has been implemented in C++. To fasten the
integration, recurrence coefficients in Eqs.~(\ref{eq.Ylm1}),
(\ref{eq.Ylm2}) and (\ref{eq.recD2}), the constant terms
$u^{l_1,m_1}_{l_2,m_2}$ (Eq.~\ref{eq.u}), 
as well as the constant values in the evaluation of the ladder operators
(Eqs.~\ref{eq.Ladder12} and \ref{eq.Ladder34}), are all calculated
and stored before the effective integration of the equations of motion
(\ref{eom}). To be homogeneous with \citetalias{Hou16}, a truncation of
the mutual potential at order $n$ means that $l_1+l_2\leq n$ ($l_1$ runs
from 0 to $n$ and $l_2$ from 0 to $n-l_1$) in the expression of
$U_{AB}$, $\vec F$, $\vec T$, and $\vec T_B$ (Sect.~\ref{sec.ladder}).

In order to compare the efficiency of the algorithm with that presented
by \citetalias{Hou16}, we integrate the binary asteroid system 1999~KW4
(66391). The polyhedral shapes of the two components are retrieved at
the URL (\url{http://echo.jpl.nasa.gov/asteroids/shapes/}). Inertia
integrals
$$
T^{l,m,n} = \int \rho x^l y^m z^n\,\dd x\dd y\dd z,
$$
with $\rho$ the (constant) density of the considered body
and where the integration is done over its whole volume, are computed
using the formulae of \citetalias{Hou16}. In particular, inertia integrals of
degree $l+m+n=0$ and 1 are used to compute the coordinates of the
barycenter which are then subtracted to those of the polyhedron
vertexes. This procedure ensures that the body-fixed frame of each
component is well centred on the barycenter as required by the
equations of motion (Eq.~\ref{eom}). The generalised products of
inertia are then converted into Stokes coefficients following
\citet{Tricarico08} but with a different normalisation factor\footnote{If we
denote by $\{C^{(T)}_{lm},S^{(T)}_{lm}\}$ Stokes coefficients defined in
\citep[Eqs. 14, 15]{Tricarico08}, those of the present paper are given
by $\{C_{lm}, S_{lm}\} = N_{lm} \{C^{(T)}_{lm},S^{(T)}_{lm}\}$ with
$$
N_{lm} = \frac{2}{1+\delta_{m,0}}\frac{(l-m)!}{(l+m)!} .
$$}. We assume the following masses: $M_A = 2.355\times10^{12}$\,kg and
$M_B = 0.135\times10^{12}$\,kg, which implies that the densities of the
two components are $1970.2$\,kg.m$^{-3}$ and $2810.5$\,kg.m$^{-3}$,
respectively.

The initial conditions are the same as in \citepalias{Hou16}: the orbital
elements are
$$
a = 2.5405\,\mathrm{km},
\quad
e = 0.01,
\quad
i=\Omega=\omega=0^\circ,
\quad
M=180^\circ\ .
$$
The initial angular speeds of the two components in their respective
body-fixed frame are
$$
\vec \Omega_A = \begin{pmatrix}
0 \\ 0 \\ 1
\end{pmatrix} 315.4^\circ.\mathrm{day}^{-1},
\qquad
\vec \Omega_B = \begin{pmatrix}
0 \\ 0 \\ 1
\end{pmatrix} 495^\circ.\mathrm{day}^{-1}\ .
$$
The 3-1-3 Euler angles defining the initial orientations of the
body-fixed reference frames relative to the initial orbit plane are
$$
\psi_A = 27.04^\circ, \quad \theta_A=10^\circ, \quad \phi_A = -83.93^\circ
\qquad
\psi_B = 0^\circ, \quad \theta_B=0^\circ, \quad \phi_B = 180^\circ\ .
$$

In order to make the most objective comparison with \citetalias{Hou16}, we
use the same RKF7(8) integrator with the same fixed time step of 200
seconds.  Tests are run on a single processor with a CPU frequency of
2.3 GHz.

\begin{figure}
\begin{center}
\includegraphics[width=0.5\linewidth]{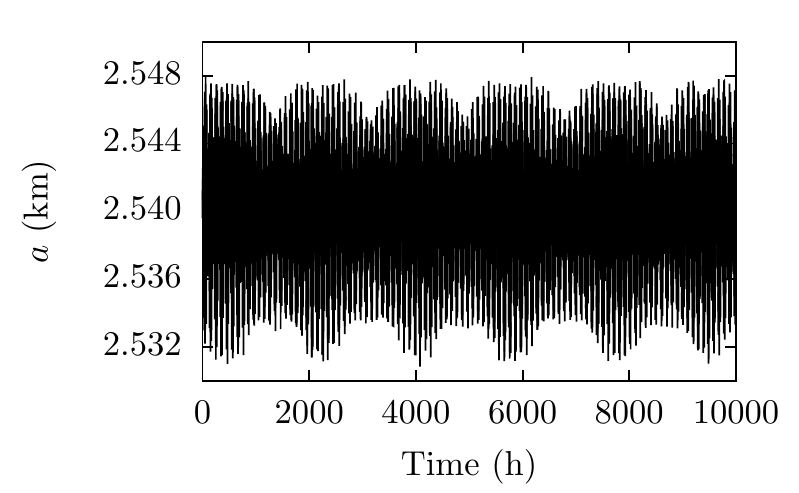}%
\includegraphics[width=0.5\linewidth]{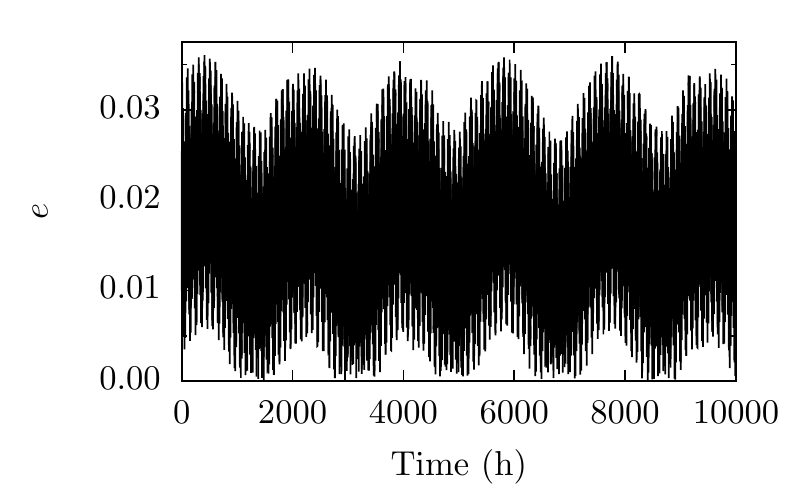}
\includegraphics[width=0.5\linewidth]{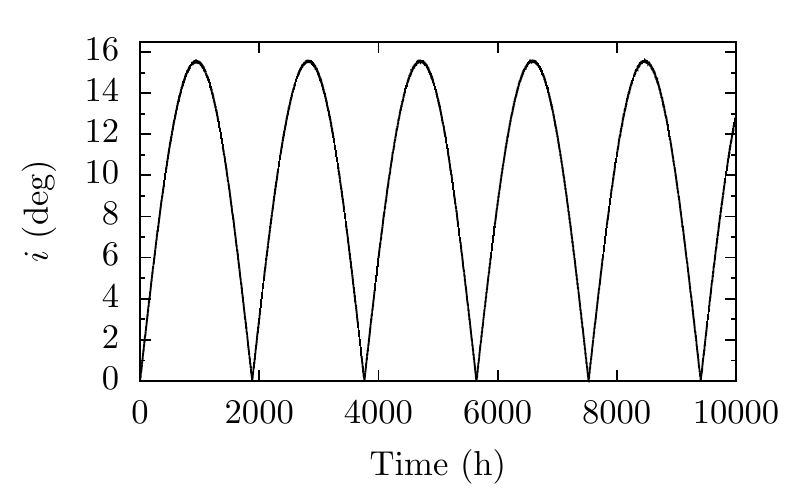}
\caption{Orbital evolution of the 1999 KW4 binary asteroid system over
$10\,000$\,h time span obtained with an RKF7(8) integrator with a fixed 
time step of 200\,s. The mutual potential is truncated at order 6. $a$
is the semi-major axis, $e$ the eccentricity, and $i$ the inclination
relative to the initial orbital plane.\label{fig.trajectory}}
\end{center}
\end{figure}

Figure~\ref{fig.trajectory} represents the orbital evolution of the system 
with respect to the initial orbital plane over $10\,000$h. The mutual
potential is truncated at order 6. As observed by \citetalias{Hou16}, the
energy stored in the rotational motion is able to strongly influence the
orbit by forcing its eccentricity and inclination to oscillate in the
intervals $0\leq e \leq 0.035$ and $0\leq i \leq 16^\circ$,
respectively. The total energy of the system is
$8.7678\times10^{10}$\,J. Numerically, its variation is mainly due to
the dissipative effect of the RKF integrator. Over the $10\,000$\,h of
integration time span, the total variation of the mechanical energy is
only $-0.3870$\,J, which is perfectly consistent with the $-0.3851$\,J
obtained by \citetalias{Hou16}. 

\begin{table}
\begin{center}
\caption{\label{tab.time}Computation time for a 200\,h integration of
the orbit using \citetalias{Hou16}'s algorithm (first row) and our
approach (second row).}
\begin{tabular*}{\textwidth}{@{\extracolsep{\fill}}l*8{r}} \hline\hline
Order & 2 & 3 & 4 & 5 & 6 & 7 & 8 & 9 \\ \hline
$t^a$ (sec) & 0.2 & 0.7 & 2.1 & 6.6  & 17.3 & 42.6 & 99.3 & 206.0 \\
$t^b$ (sec) & 0.2 & 0.3 & 0.4 & 0.6  &  1.0 &  1.7 &  2.5 &   3.1 
\\ \hline
\end{tabular*} \\
{\scriptsize 
$t^a$ computation time using \citeauthor{Hou16}'s algorithm
on a single 2.9\,GHz CPU frequency processor. \\
$t^b$ computation time using the present algorithm on a single 2.3 GHz CPU
frequency processor.
}
\end{center}
\end{table}

A comparison of the integration times between our approach and the
algorithm of \citetalias{Hou16} is presented in Table~\ref{tab.time}. The
system is integrated over 20\,000\,h with a fixed time step of 200\,s,
and then the computation time is divided by 100 for a direct comparison
with \citetalias{Hou16} who did the integration of the orbit over 200\,h.
Absolute CPU times highly depend on the compiler and the optimisation
options. They should thus be compared with care. Nevertheless, both
methods spend the same amount of computation time when the potential is
truncated at the lowest order $n=2$. We can thus assume that the other
columns of Tab.~\ref{tab.time} objectively reflect the efficiency of
the algorithms. Moreover, as the expansion order increases, the
computation time growth differs significantly between the two methods.
Ours proves to be much faster. At the 6$^\mathrm{th}$ order of
truncation, we already gain a factor 17 in speed, and at order 9, our
approach only needs 3 seconds instead of the 3.4 minutes required by the
algorithm of \citetalias{Hou16}.

\section{Conclusion}
\label{sec.conclusion}
The decomposition of the mutual potential of two rigid bodies with
arbitrary shape into spherical harmonics has sometimes been discarded
because of the misconception that this formalism would involve
trigonometric functions. However spherical harmonics do have expressions
in terms of Cartesian coordinates. Moreover, not only recurrence
relations allow to evaluate them efficiently, but the force and the
orbital torque are easily computed by application of the gradient and
the angular momentum operators, respectively. By consequence, if we
restrain ourselves to the orbital part of the equations of motion only,
our approach must be as efficient as a polynomial decomposition
of the potential in Cartesian coordinates.

Nevertheless, numerical tests show that our algorithm is faster. The
reason resides in the way rotations are handled. Here, we use the
irreducible representation of the group SO(3) acting on the set of
functions defined on the sphere, viz. Wigner D-matrices. These matrices
have two advantages: they can be efficiently evaluated with recurrence
relations and the torque is simply obtained by application of the spin
operator. Combining spherical harmonics and Wigner D-matrices, we obtain
an algorithm for which forces and torques are calculated as rapidly as
the mutual potential.

In this work, in order to avoid singularities, rotations are parametrised
using Cayley-Klein's formalism whose parameters are strictly equivalent
to quaternions. We make this choice because it is well adapted to the
computation of Wigner D-matrices. Because this representation in not
common in the celestial mechanics field, we briefly summarise the
relation between these parameters and the standard 3-1-3 Euler angles.
More properties can be found in textbooks on quantum theory of
angular momentum \citep[e.g.,][]{Varshalovich88}.

As a concluding remark, not only the present formalism allows to design
a fast integrator for the full two rigid body problem, but it also
reveals the underlying structure which is controlled by the group
of rotations. This structure is often hidden in other approaches. We
expect that this will allow to draw a more comprehensive view of
the analytical problem.

\bibliographystyle{spbasic}      
\bibliography{2body}   

\end{document}